\documentclass{article}

\usepackage[nonatbib, final]{neurips_2022_ml4ps}

\usepackage[utf8]{inputenc} % allow utf-8 input
\usepackage[T1]{fontenc}    % use 8-bit T1 fonts
\usepackage{hyperref}       % hyperlinks
\usepackage{url}            % simple URL typesetting
\usepackage{booktabs}       % professional-quality tables
\usepackage{amsfonts}       % blackboard math symbols
\usepackage{nicefrac}       % compact symbols for 1/2, etc.
\usepackage{microtype}      % microtypography
\usepackage{xcolor}         % colors
\usepackage{graphicx}
\usepackage{amsmath}
\usepackage{float}

\usepackage[numbers]{natbib}
\title{Recovering Galaxy Cluster Convergence from Lensed CMB with Generative Adversarial Networks}

\author{%
  Liam Parker \\
  Princeton University\\
  Princeton, NJ\\
  \texttt{lhparker@princeton.edu} \\
  \And
  Dongwon Han \\
  University of Cambridge\\
  Cambridge, UK \\
  \texttt{dh689@cam.ac.uk} \\
  \And
  Pablo Lemos Portela\\
  Mila\\
  Montreal, Quebec\\
  \texttt{plemos91@gmail.com}\\
  \And
  Shirley Ho \\
  CCA, Flatiron Institute\\
  New York, NY \\
  \texttt{shirleyho@flatironinstitute.org} \\
}

\begin{document}

\maketitle

\begin{abstract}
  We present a new method which leverages conditional Generative Adversarial Networks (cGAN) to reconstruct galaxy cluster convergence, $\kappa$, from lensed CMB temperature maps. Our model is constructed to emphasize structure and high-frequency correctness relative to the Residual U-Net approach presented by Caldeira, et. al. (2019). Ultimately, we demonstrate that while both models perform similarly in the no-noise regime (as well as after random off-centering of the cluster center), cGAN outperforms ResUNet when processing CMB maps noised with $5\mu K/arcmin$ white noise or astrophysical foregrounds (tSZ and kSZ); this out-performance is especially pronounced at high $\ell$, which is exactly the regime in which the ResUNet under-performs traditional methods. 
\end{abstract}

\section{Introduction}
The universe's oldest light, the cosmic microwave background (CMB), encodes in its temperature and polarization a wealth of information. While the study of the CMB has a long and multi-faceted history, the study of the gravitational effects imprinted on the CMB at later stages of the universe has captured the attention of modern cosmology. 

Specifically, the study of gravitational lensing of the CMB by galaxy clusters has enabled the reconstruction of the gravitational potential field, $\phi$, around these clusters. For one, this field contains information about the spatial distribution of these clusters, which enables insights into the parameters that govern their formation, such as dark energy and massive neutrino properties \citep{lewis2006weak}. Moreover, a variety of cosmological parameters, such as $\Omega_m$, the mass density of the universe, can be constrained by recovering the total mass of these clusters from $\phi$ \citep{bocquet2019cluster}. However, with recent and upcoming CMB surveys - e.g. AdvancedACTPol \citep{thornton2016atacama} and Simons Observatory \citep{ade2019simons} -  expected to amass lensed CMB measurements at unprecedentedly high signal-to-noise ratios, discoveries tied to the study of CMB lensing are likely to only become more significant. 

Currently, the most prevalent method for reconstructing $\phi$ from lensed CMB is the quadratic estimator (QE), an estimator formed from quadratic combinations of data \citep{hu2002mass}. However, QE is shown to be sub-optimal for low-noise polarization data due to the lensing itself \citep{yoo2008improved}, as well as for low-noise temperature data due to the cosmic variance of the background CMB gradient \citep{hadzhiyska2019improving} \cite{hirata2003analyzing}, and thus not suited for the higher signal-to-noise ratios promised by this novel generation of CMB surveys. As such, a variety of alternatives have been proposed, including a gradient-inversion technique \cite{hadzhiyska2019improving}, a maximum-likelihood-estimator \cite{raghunathan2017measuring}, and a hierarchical Bayesian inference method \cite{muse}.

Machine learning (ML)-backed methods also present an attractive alternative to QE. Indeed, Caldeira, et. al. \cite{caldeira2019deepcmb} use  a Residual U-Net (ResUNet) to recover $\kappa$ maps (the dimensionless surface-mass density along the line of sight) around galaxy clusters from input lensed CMB with higher signal-to-noise ratios than QE over a broad range of angular scales in the low-noise regime. However, the quality of ResUNet's predictions materially degrades in noisier conditions, as well as at higher angular scales, where it underperforms QE. 

One disadvantage of using the ResUNet method is its reliance on a static loss function (such as L1/L2) during network optimization. While this static loss function has been shown to capture low-frequency components, it essentially formulates the image-to-image translation problem as a per-pixel regression problem, thereby ignoring dependence between pixels in the output space and often leading to a loss of sharpness and structure \citep{larsen2016autoencoding}. 

In this work, we aim to overcome this disadvantage by optimizing the ResUNet with a trainable loss function (discriminator) in conjunction with an L1 loss, effectively transforming the ResUNet into a modified Pix2Pix conditional generative adversarial network (cGAN) \cite{cgan}. This architecture is particularly adept at extrapolating structure and high-frequency components in image-to-image translation tasks while maintaining low-frequency correctness. We train both our cGAN and a ResUNet to recover $\kappa$ maps from lensed CMB temperature maps around galaxy clusters under four conditions: no noise, astrophysical (tSZ and kSZ) foreground noise, $5 \mu K/arcmin$ white noise (which mimics instrumentation noise), and random off-centering (which mimics the fact that the clusters centers will not always be perfectly centered). We demonstrate that our cGAN outperforms ResUNet under all four conditions, and that this out-performance becomes especially pronounced in the noisier regimes and at high $\ell$ \footnote{As the present paper focuses on reconstruction in the high-$\ell$ regime, and because of the fact that QE is computationally prohibitively expensive, we focus on comparing our cGAN architecture exclusively to ResUNet.}.

\section{Data}
We employ the Websky Extra-galactic CMB Simulations \citep{websky} to model the lensed CMB temperature anisotropies, tSZ and kSZ effects, and $\kappa$ maps around corresponding galaxy clusters. These simulations are tailored to the upcoming ground-based CMB surveys such as the Simons Observatory and Advanaced ACTPol \citep{websky}, rendering them ideal for our purposes \footnote{The following cosmological parameters are used in the Websky simulations: $\Omega_m = 0.31, \Omega_b = 0.049, \Omega_c = 0.261, H_0 = 100*0.68, n_s = 0.965, \tau = 0.0943$ \cite{websky}.}. We select $50,000$ clusters matching the following criterion: $M_{200m} \in [10^{13},5*10^{14}] M_{\odot}$ and $z \in [0.47,0.6]$, and cut out the lensed CMB temperature, $\kappa$, tSZ, and kSZ maps in $128 \times 128$ arcmin squares around the cluster center. Additionally, we create random $5\mu K/arcmin$ $128\times128$ white noise maps. Notably, all maps are projected into 2D euclidean space, as our models are only capable of handling such formats, using Orphics \cite{orphics}. From these maps, we compile three feature datasets, $X_{CMB}$, $X_{CMB+tSZ+kSZ}$, and $X_{CMB+5\mu K}$, and use the $\kappa$ maps as our target dataset. Additionally, we repeat this process for the pure CMB and $\kappa$ maps with random off-centering in both RA and Dec directions by a Gaussian with mean 0 and 1 arcmin variance, resulting in one additional feature/target pair, $\{X_{oc}, \kappa_{oc}\}$. All datasets are thus size $(50000, 128, 128, 1)$, and we split each using a 80:10:10 split.

\section{Method}
Our cGAN is made up of two main components: $G$, the generator, and $D$, the discriminator. Both $G$ and $D$ are convolutional neural networks (CNNs), which have demonstrated unparalleled power in dealing with image-data \citep{aloysius2017review}; moreover, $G$ is a Residual U-Net, a type of CNN which uses an archetypal encoder-decoder scheme that has proven adept at a wide array of image-to-image translation tasks \citep{zhang2018road}. $G$ learns to map the observed CMB temperature map, $X$, to a predicted $\kappa_{pred}$ map, $G : X \rightarrow \kappa_{pred}$, while $D$ takes as input the concatenated generator-predicted $\kappa_{pred}$ and ground-truth $\kappa$ maps, and predicts a grid of the likelihood $[0, 1]$ that each $70\times70$ patch in $\kappa_{pred}$ is real, $D:\{\kappa_{pred}, \kappa\} \rightarrow D_{output}$. We model $G$ after the ResUNet proposed by Caldeira, et. al. \cite{caldeira2019deepcmb}, albeit with some modifications, and $D$ after the convolutional PatchGAN architecture proposed by \cite{cgan}. Ultimately, $G$ is trained to produce $\kappa_{pred}$ maps as similar as possible to the ground truth $\kappa$ map, while $D$ is trained to discriminate between the "fake" $\kappa_{pred}$ and the ground-truth $\kappa$. 

\begin{figure}
    \centering
    \includegraphics[scale=0.225]{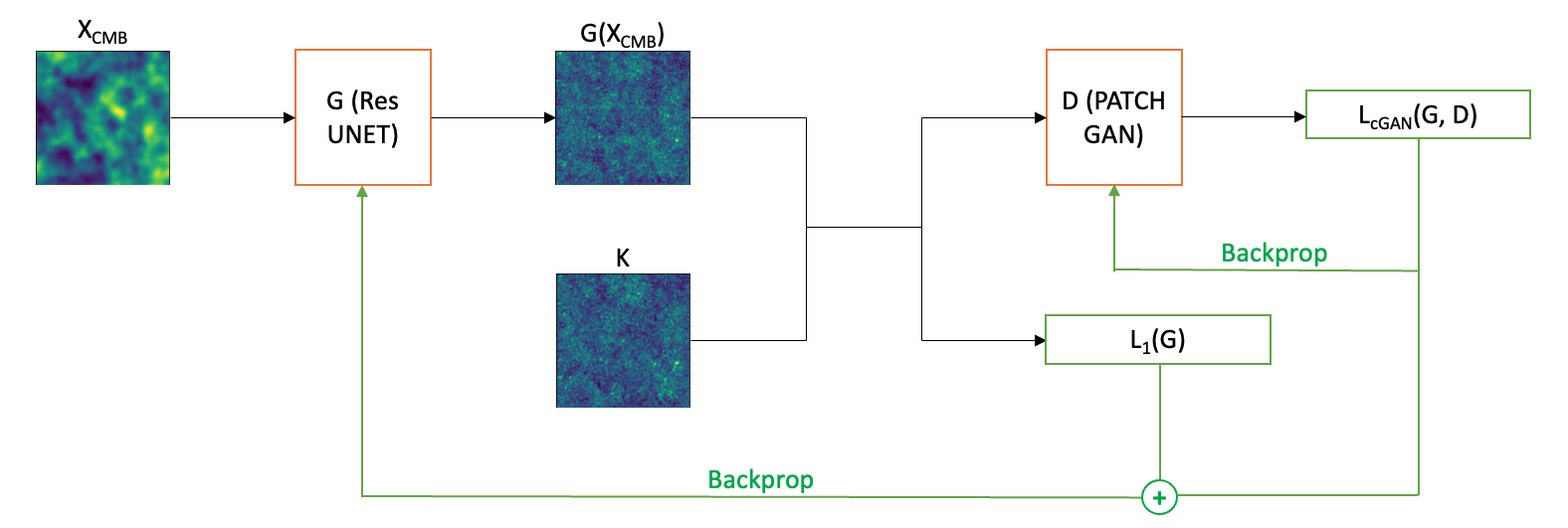}
    \caption{Structure of the cGAN. As illustrated, the generator, $G$, based on the ResUNet architecture takes as input $X_{CMB}$ and produces a predicted $\kappa$ map. These are fed into the discriminator, $D$, to produce the first loss function, $L_{cGAN}(G,D)$ used to update the parameters of both $G$ and $D$. An $L_1$ loss is also included, as it has been shown to emphasize low-frequency correctness.}
    \label{fig:my_label}
\end{figure}

$G$ is composed of an encoder, a bottleneck, and decoder. The encoder is made of six encoding blocks, each of which consists of a down-sampling convolutional layer followed by a simple convolutional layer. The number of filters for each block increases as $32*2^{n}$, where $n$ is the index of the encoder block from $n \in \{0,5\}$. The decoder follows the same structure, however the first convolutional layer of each decoder block is a 2D transpose convolutional layer, and the number of filters goes as $32*2^{5-n}$. Residual connections are established between encoder and decoder blocks with same $n$ to allow low-level feature information to flow more easily through the network. The output of the decoder block is fed into a final 2D transpose convolutional layer after which $tanh$ activation is applied \footnote{For consistency, we use this same network architecture and parameters for the ResUNet to which we compare our cGAN.}. $D$ is composed of four encoding convolutional layers with filter size $64*2^n$, followed by a simple convolutional layer with filter size $512$, a final convolutional layer with filter size $1$, and a $sigmoid$ activation function. Batch normalization is applied to all convolutional layers in $G$ and $D$ except the first, and $Leaky ReLU$ activation functions follow each convolutional layer except for the last. Additionally, the first two decoding blocks in $G$ have $dropout=0.5$ applied. The models are constructed using the Keras interface of the Tensorflow API \citep{abadi2016tensorflow}, and we use the general GAN update structure laid out by Brownlee \cite{brownlee_2021}. 

We train both our cGAN and the ResUNet three separate times on the $40,000$ cluster $\{X_{CMB}, \kappa\}, \{X_{CMB+tSZ+kSZ}, \kappa\}$, $\{X_{CMB+5\mu K}, \kappa\}$, and $\{X_{oc}, \kappa_{oc}\}$ training datasets. Notably, cGAN is trained to minimize both the discriminator loss and an L1 loss; more specifications on optimization are provided in the Appendix. We train both models for 100 epochs using mini-batches of 32 samples on a single NVIDIA Tesla P100 GPU with 16GB of Graphic RAM, which takes 3.8 hours for the ResUNet and 8.1 hours for the cGAN.

\section{Results}
\begin{figure}
    \centering
    \includegraphics[scale=0.3]{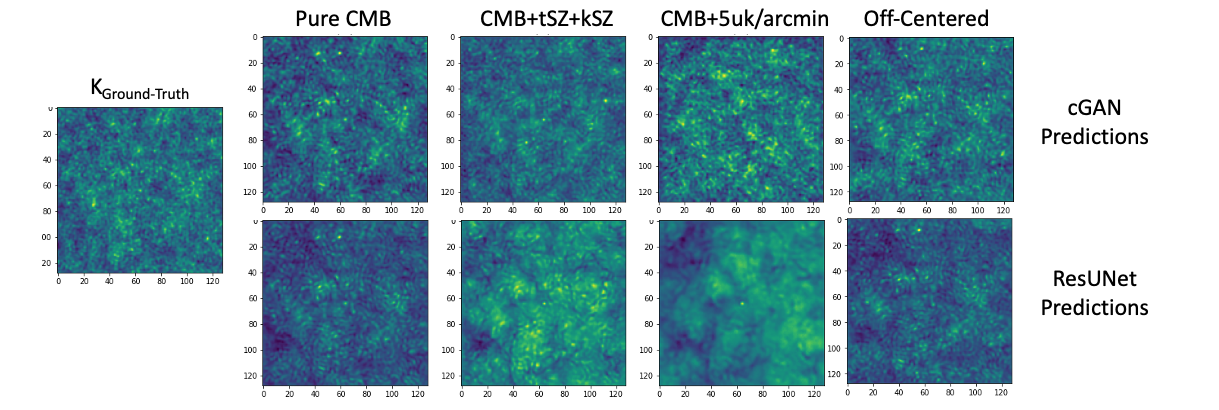}
    \caption{Visualization of sample predicted $\kappa_{pred}$ maps from the cGAN and ResUNet for a random cluster in the test dataset. The predictions are made for CMB temperature maps with either no noise, astrophysical foreground noise (tSZ+kSZ), $5\mu K/arcmin$ white noise, or random off-centering according to a Gaussian with $1 arcmin$ variance.}
    \label{fig:visuals}
\end{figure}

\begin{figure}
    \centering
    \includegraphics[scale=0.4]{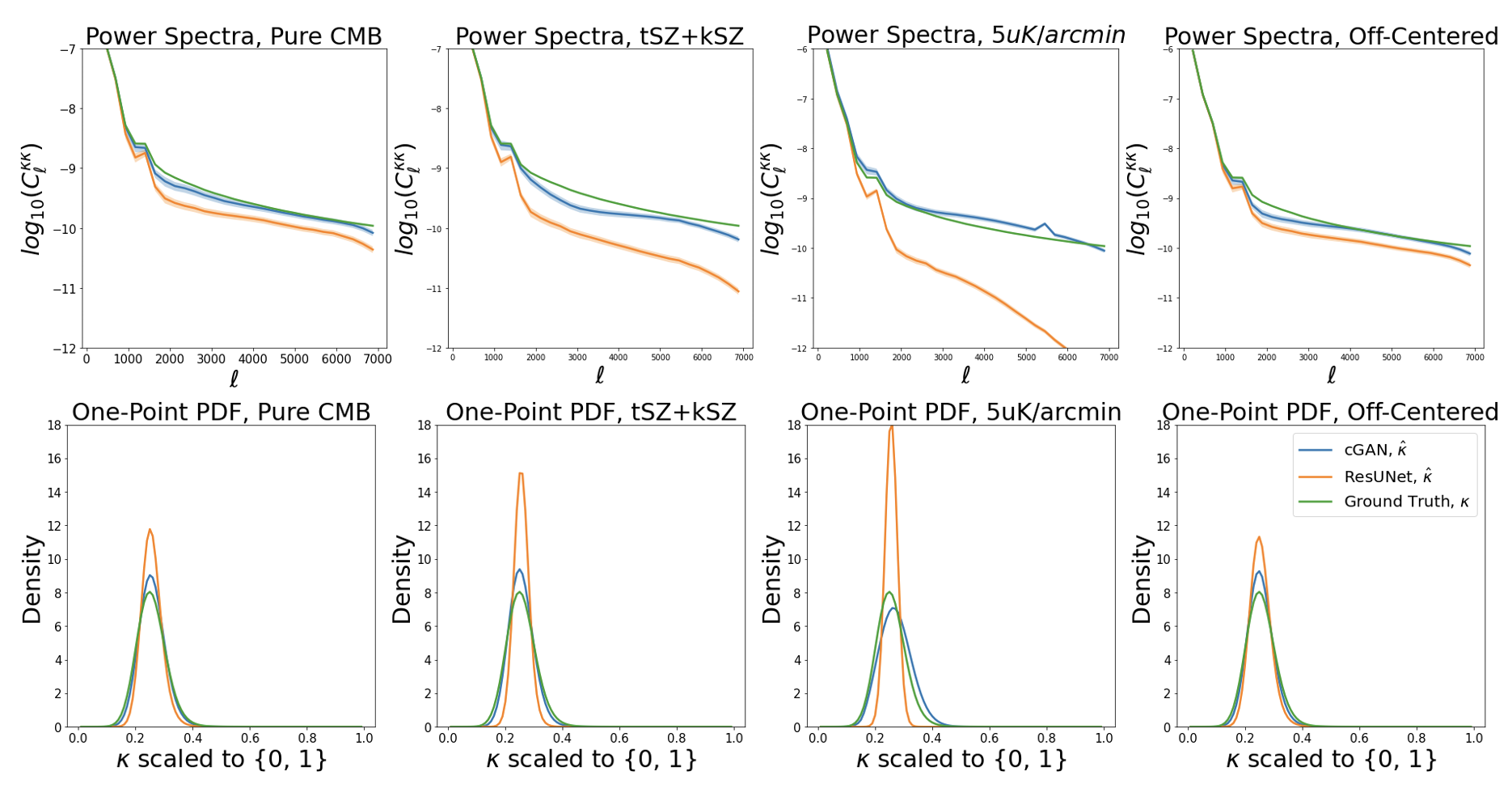}
    \caption{The mean over the 5000-cluster test dataset of the one-point PDF/power spectra of the cGAN and ResUNet predicted $\kappa_{pred}$ maps under various noise conditions. For the power spectra, we also calculate the standard deviation in predicted power spectra over the test dataset using the bootstrap method, and include these standard deviations in as the shaded region in the graphs (the average standard deviation per cluster is on the order of magnitude of 10\% for all predictions).}
    \label{fig:quant}
\end{figure}

All results are generated over the held-out $5000$-cluster test dataset. Figure \ref{fig:visuals} provides a sample visualization of ResUNet and cGAN predictions under various noise conditions for a random test cluster. From visual inspection, it appears as if cGAN captures significantly more information than ResUNet under all noise conditions; this is especially apparent in the noised (astrophysical foreground and $5\mu K/arcmin$) regimes, where cGAN continues to recover the majority of structural information, whereas significant blurring and/or loss of structure has occurred in the ResUNet predictions. In order to quantitatively test the performance of cGAN, we compare the power spectrum (generated using the Orphics package \cite{orphics}) and one-point PDF of the predictions of cGAN and ResUNET to the ground-truth $\kappa$ maps in Figure \ref{fig:quant}.

In observing the mean power spectra, it is clear that the ResUNet power spectrum is able to mimic that of the ground-truth relatively well in the noiseless regime, as well as after random off-centering. However, it materially diverges from the ground-truth power spectrum under both astrophysical foreground and $5\mu K/arcmin$ noise regimes; this divergence is especially pronounced at high $\ell$. Conversely, the cGAN power spectrum is able to stay faithful to the ground-truth under all four regimes, and does not materially degrade even at high $\ell$: no noticeable degradation is visible until around $\ell \approx 6500$. The relative strength of the cGAN is further emphasized in the average one-point PDFs, in which ResUNet's one-point PDF materially diverges from the ground-truth $\kappa$ map under the noised regimes while cGAN's does not. Ultimately, cGAN's successes at high $\ell$, highlight the relative ability of the discriminator architecture at emphasizing small-scale structure in the predicted $\kappa$ maps.

\section{Conclusion}
In the present paper, we demonstrate that the inclusion of a discriminator in the optimization of a Residual U-Net can materially improve its performance in recovering galaxy cluster convergence from lensed CMB temperature maps. Specifically, across both visualizations of predicted $\kappa$ maps, as well as the power spectra and one-point PDFs of these $\kappa$ maps, we show that the disciminator-enhanced network (cgAN) noticeably outperforms ResUNet under a variety of noise conditions. Moreover, we demonstrate that this out-performance becomes especially pronounced in noisy regimes (such as instrumentation or astrophysical foreground noise), as well as at high $\ell$. This out-performance at high $\ell$ is particularly encouraging, as small-scale features are challenging to recover using traditional QE methods. In future work, it will be valuable to explore 1) cGAN's performance under a wider variety of noise conditions, 2) cGAN's relative performance with additional loss functions (such as a Fourier-space loss), and 3) alternative GAN structures (such as Wasserstein GAN \cite{wgan}). 

\section{Broader Impact}
We employ a generative adversarial network (GAN) - a popular machine learning model - in an astrophysical context. GANs are already present in myriad social applications, and while most use cases are benign, they have been maliciously employed across multiple social media platforms, from generating fake Facebook accounts to conducting personation attacks on targeted subjects \cite{cnn_facebook}. The applications of such algorithms to astrophysics however has been quite limited. Nonetheless, in applying GANs in these contexts, we can better understand where and why they fail, which can help improve our ability to spot maliciously employed GANs more broadly.

\begin{ack}
The contributions to the paper are as follows. \textbf{Parker}; performed all neural network computation work; innovated NN architecture; performed NN diagnostic analysis; wrote up the paper and corresponding poster; initiated problem concept. \textbf{Han}; constructed data pipeline from Websky; guided ResUNet construction; designed NN diagnostic analysis; initiated problem concept. \textbf{Portela}; consulted on paper edits; helped clarify sticky physics questions. \textbf{Ho}; designed uncertainty calculations; suggested one-point PDF analysis; guided paper through the final steps before submission.

Moreover, we are extremely thankful and indebted to Professor Suzanne Staggs for the invaluable time, advice, and effort that she has graciously provided in guiding us during the completion of this fascinating project. 

\end{ack}

\bibliography{bib}

%%%%%%%%%%%%%%%%%%%%%%%%%%%%%%%%%%%%%%%%%%%%%%%%%%%%%%%%%%%%
\section*{Checklist}

%%% BEGIN INSTRUCTIONS %%%
The checklist follows the references.  Please
read the checklist guidelines carefully for information on how to answer these
questions.  For each question, change the default \answerTODO{} to \answerYes{},
\answerNo{}, or \answerNA{}.  You are strongly encouraged to include a {\bf
justification to your answer}, either by referencing the appropriate section of
your paper or providing a brief inline description.  For example:
\begin{itemize}
  \item Did you include the license to the code and datasets? \answerYes{See Section 2.}
  \item Did you include the license to the code and datasets? \answerNo{The code and the data are proprietary.}
  \item Did you include the license to the code and datasets? \answerNA{}
\end{itemize}
Please do not modify the questions and only use the provided macros for your
answers.  Note that the Checklist section does not count towards the page
limit.  In your paper, please delete this instructions block and only keep the
Checklist section heading above along with the questions/answers below.
%%% END INSTRUCTIONS %%%

\begin{enumerate}

\item For all authors...
\begin{enumerate}
  \item Do the main claims made in the abstract and introduction accurately reflect the paper's contributions and scope?
    \answerYes{}{}
  \item Did you describe the limitations of your work?
    \answerYes{}
  \item Did you discuss any potential negative societal impacts of your work?
    \answerYes{}
  \item Have you read the ethics review guidelines and ensured that your paper conforms to them?
    \answerYes{}
\end{enumerate}

\item If you are including theoretical results...
\begin{enumerate}
  \item Did you state the full set of assumptions of all theoretical results?
    \answerNA{}
        \item Did you include complete proofs of all theoretical results?
    \answerNA{}{}
\end{enumerate}

\item If you ran experiments...
\begin{enumerate}
  \item Did you include the code, data, and instructions needed to reproduce the main experimental results (either in the supplemental material or as a URL)?
    \answerNo{} The code will be available upon acceptance. 
  \item Did you specify all the training details (e.g., data splits, hyperparameters, how they were chosen)?
    \answerYes{} With additional training details available in the Appendix. 
        \item Did you report error bars (e.g., with respect to the random seed after running experiments multiple times)?
    \answerYes{} 
        \item Did you include the total amount of compute and the type of resources used (e.g., type of GPUs, internal cluster, or cloud provider)?
    \answerYes{}
\end{enumerate}

\item If you are using existing assets (e.g., code, data, models) or curating/releasing new assets...
\begin{enumerate}
  \item If your work uses existing assets, did you cite the creators?
    \answerYes{} All open-source code and datasets have been cited in the body of the paper.
  \item Did you mention the license of the assets?
    \answerNA{}
  \item Did you include any new assets either in the supplemental material or as a URL?
    \answerNA{}
  \item Did you discuss whether and how consent was obtained from people whose data you're using/curating?
    \answerNA{}
  \item Did you discuss whether the data you are using/curating contains personally identifiable information or offensive content?
    \answerNA{}
\end{enumerate}

\item If you used crowdsourcing or conducted research with human subjects...
\begin{enumerate}
  \item Did you include the full text of instructions given to participants and screenshots, if applicable?
    \answerNA{}
  \item Did you describe any potential participant risks, with links to Institutional Review Board (IRB) approvals, if applicable?
    \answerNA{}
  \item Did you include the estimated hourly wage paid to participants and the total amount spent on participant compensation?
    \answerNA{}
\end{enumerate}

\end{enumerate}

\appendix
\section{Network Optimization}
The cGAN network is trained by optimizing the GAN loss function from Isola, et. al. \cite{cgan}, given by
\begin{align}
    L_{cGAN}(G,D) = \mathbb{E}_{\kappa}[log D(\kappa)] + \mathbb{E}_{X_{CMB}}[log(1-D(G(X_{CMB}))].
\end{align}
Because this loss function is driven by $D$, and $D$ is constructed to emphasize localized structure recovery and high-frequency correctness, we include an $L_1$ loss as well in order to promote low-frequency correctness in the output space, 
\begin{align}
    L_1(G) = \mathbb{E}_{X_{CMB}, \kappa}[\lVert \kappa - G(X_{CMB}) \rVert]_{L_1}.
\end{align}
Thus, the total network loss is the weighted sum of $L_{cGAN}$ and $L_1$ , where we weight $L_1$ by a factor of 100 relative to $L_{cGAN}$ as suggested in Isola, et. al. \cite{cgan}. This gives a composite objective function
\begin{align}
G^*_{cGAN} = arg \underset{G}{min}\underset{D}{max} L_{cGAN}(G,D) + 100L_{L1}(G),
\end{align}
which is minimized relative to $D$ (as $D$ should maintain a maximum distance between real and fake in every iteration) and maximized relative to $G$.

During cGAN training, we alternate between one update step on $D$ and one update step on $G$. Specifically, we (1) generate a batch of fake samples, $G(X_{CMB}) \rightarrow \kappa_{pred}$, corresponding to a real batch of samples $\kappa$, (2) update the discriminator using the batch of real $\kappa$, (3) update the discriminator using the batch of generated $\kappa_{pred}$, (4) update the generator using the composite loss. Importantly, we divide the objective by 2 while optimizing $D$, slowing down its learning rate relative to $G$, as otherwise $D$ has been found to outpace $G$ \citep{cgan}. Our training is run on the Adam optimizer \citep{adam}, with a learning rate of $2 \times 10^{-4}$ and $\beta_1 = 0.5, \beta_2 = 0.999$. Additionally, for the ResUNet, we use a learning rate scheduler as well, which reduces the learning rate by a half every time the training loss has not improved for five consecutive epochs.

%%%%%%%%%%%%%%%%%%%%%%%%%%%%%%%%%%%%%%%%%%%%%%%%%%%%%%%%%%%%

\end{document}